\documentclass[superscriptaddress, twocolumn]{revtex4-1}

\usepackage{amssymb,amsmath}
\usepackage{latexsym}
\usepackage{graphicx}
\usepackage{caption}
\usepackage{subcaption}
\usepackage{enumerate}

\usepackage{color}
\usepackage{graphics}
\usepackage{epsfig}
\usepackage{bm}

\newcounter{mnotecount}[section]

\begin{document}

\newcommand{\dR}{\mathbb R}
\newcommand{\dC}{\mathbb C}
\newcommand{\dZ}{\mathbb Z}
\newcommand{\id}{\mathbb I}
\newtheorem{theorem}{Theorem}
\newcommand{\ud}{\mathrm{d}}
\newcommand{\mfn}{\mathfrak{n}}

\author{Przemys{\l}aw Ma{\l}kiewicz}
\email{Przemyslaw.Malkiewicz@ncbj.gov.pl}

\affiliation{National Centre for Nuclear Research,  00-681
Warszawa, Poland}
\author{Artur Miroszewski}
\email{Artur.Miroszewski@ncbj.gov.pl}

\affiliation{National Centre for Nuclear Research,  00-681
Warszawa, Poland}

\date{\today}

\title[]{Internal clock formulation of quantum mechanics}

\begin{abstract}
The basic tenet of the present work is the assumption of the lack of external and fixed time in the Universe. This assumption is best embodied by general relativity, which replaces the fixed space-time structure with the gravitational field, which is subject to dynamics. The lack of time does not imply the lack of evolution but rather brings to the forefront the role of internal clocks which are some largely arbitrary internal degrees of freedom with respect to which the evolution of timeless systems can be described.  We take this idea seriously and try to understand what it implies for quantum mechanics when the fixed external time is replaced by an arbitrary internal clock. We put the issue in a solid, mathematically rigorous framework. We find that the dynamical interpretation of a quantum state of a timeless system depends on the employed internal clock. In particular, we find that the continuous spectra of well-known dynamical observables like the position of a free particle on the real line may turn discrete if measured in unusual clocks. We discuss the meaning of our result for attempts at quantization of global gravitational degrees of freedom.
\end{abstract}

\pacs{98.80.Qc} \maketitle

\section{Introduction}

\begin{figure}[t]
\centering
\includegraphics[width=0.4\textwidth]{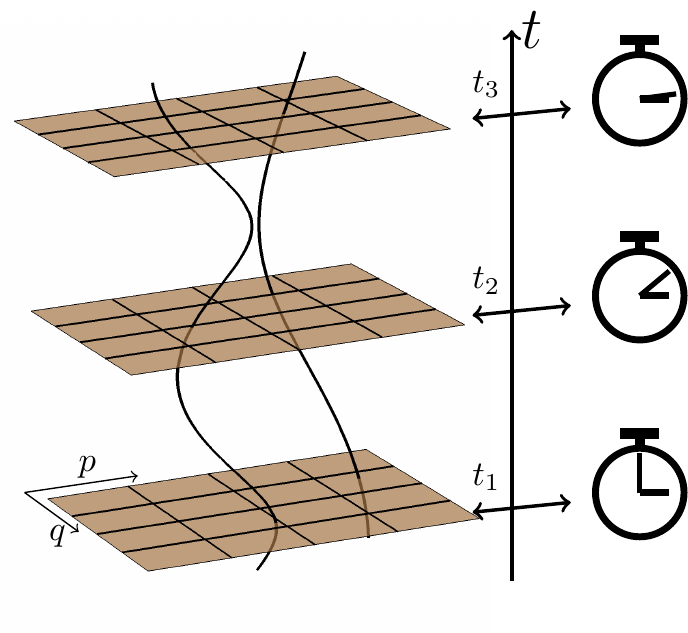}
\caption{The evolution of a given subsystem is always expressed in relation to another subsystem that is represented in the figure by a clock. Time is only an auxiliary parameter, which is used to formally isolate a given subsystem from the rest of the system, which can be then neglected. }
\label{1}
\end{figure}

The purpose of the present paper is to develop a description of mechanics in terms of internal clocks. Let us first explain why such a formulation of mechanics is desirable. In nonrelativistic  classical and quantum mechanics, there is no observable related to time. Time is assumed to be a fixed, external entity, which is not measurable. Therefore, in practice, we describe the change of a system in relation to another one. The latter we shall call the ``internal clock" (see Fig. \ref{1}). Some internal clocks are better than others. In principle, they should evolve monotonically on the relevant time scales. Furthermore, they should not interact significantly with the observed subsystems; otherwise, the observed dynamics will be difficult to understand. For example, in the context of astronomy, the ``astronomical time", given by the Earth's rotation, can serve as an example of the imperfect clock.  Around the turn of the 20th century it was finally replaced by the so-called ephemeris time which involves the motions of the Moon, the Earth and the Sun. This switch enabled us to predict more accurately positions of celestial bodies and especially of the Moon \cite{ED}. \\

%\subsection*{General relativity and internal clocks}

\begin{figure}[t]
\centering
\captionbox
{The constraint surface $C=0$ embedded in an extended phase space can be identified with a contact manifold made of a lower-dimensional phase space by choosing an internal clock. \label{int}}
{\includegraphics[width=0.5\textwidth]{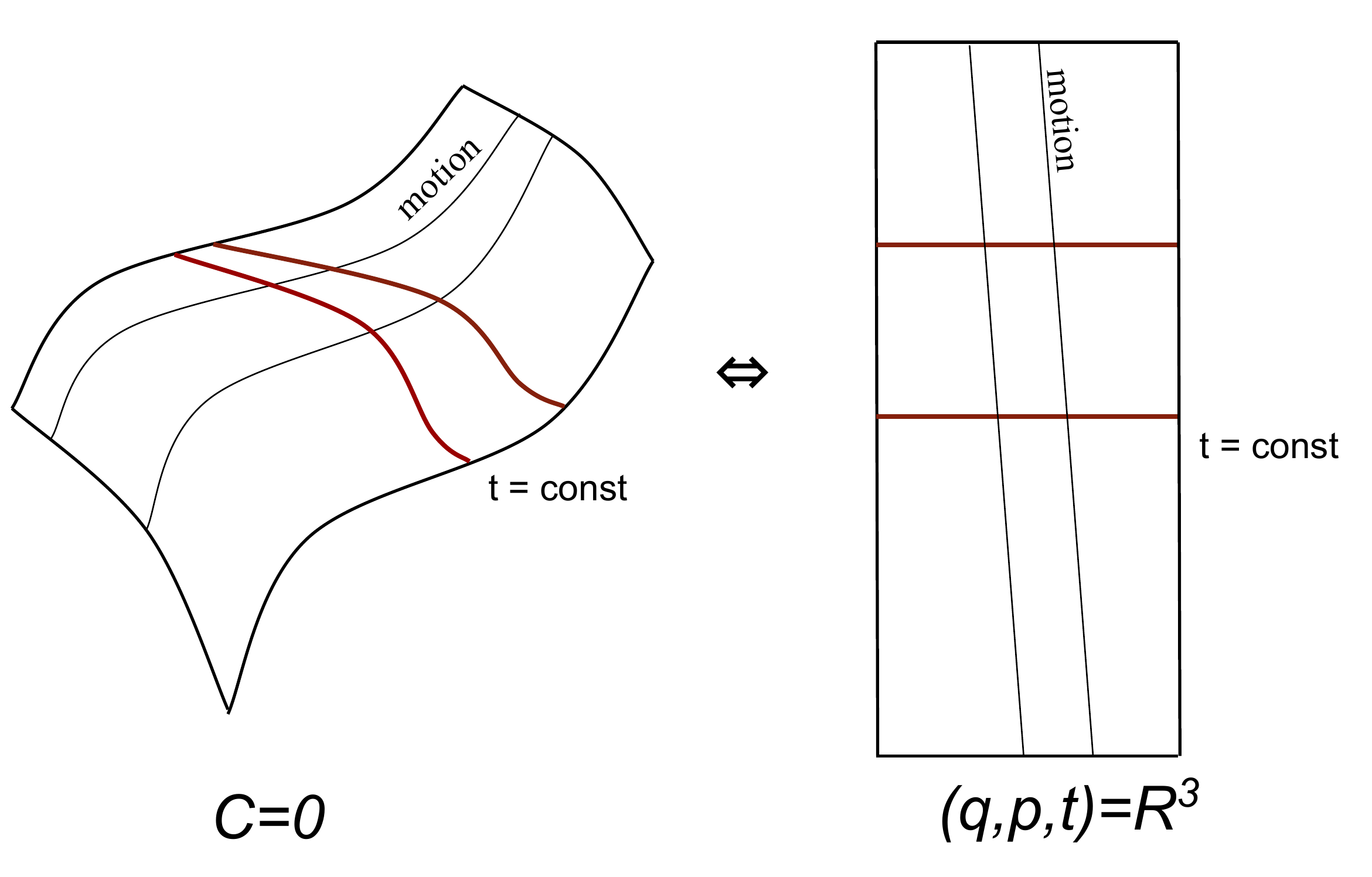}}

\end{figure}

The key motivation for the present work follows from general relativity. Since the space-time is a dynamical entity itself, there is neither a predetermined time nor a predetermined causal structure. Rather, the gravitational and material fields evolve combined together and only internal degrees of freedom can provide a physical measure of their evolution. This, as we will show, has significant consequences with regard to the formulation of quantum mechanics. It is worth it to point out that some important issues related to time already appear in quantum mechanics based on an absolute time and they are broadly discussed in the reviews \cite{R1,R2}. There might be some deep connection between these issues and our result but it is beyond the scope of the present paper.\\
The phase space formalism of general relativity \cite{adm} involves a Hamiltonian, which is constrained to vanish. The Hamiltonian constraint, say $C$, plays two roles in this formalism: (i) generating the dynamics and (ii) constraining the space of physically admissible states, 
\begin{align}\label{tau}\begin{split}
\frac{\ud}{\ud \tau}O(q_i,p_i)=\{O(q_i,p_i),C(q_i,p_i)\},\\ C(q_i,p_i)=0.\end{split}
\end{align}
Such a formalism admits the so-called time-reparametrization invariance as the constraint function $C$ may be multiplied by any nonvanishing function $N(q_i,p_i)$ leading to
\begin{align}\begin{split}
\frac{\ud}{\ud \tau'}O(q_i,p_i)=\{O(q_i,p_i),N\cdot C(q_i,p_i)\},\\ N\cdot C(q_i,p_i)=0,\end{split}
\end{align}
which is a set of equations equivalent to Eqs (\ref{tau}) by the virtue of the reparameterization of the time parameter along each dynamical trajectory,
\begin{align}
N\ud \tau'=\ud \tau.
\end{align}
The time-reparametrization invariance indicates that $\tau$ and $\tau'$ are in fact auxiliary parameters devoid of the absolute status of the Newtonian time. The Hamiltonian constraint formalism can be brought to the ordinary canonical formalism upon identifying the odd-dimensional constraint surface $C=0$ embedded in  a higher-dimensional phase space with a contact manifold made of a lower-dimensional phase space and a time manifold. This procedure is called the reduced phase space approach. To make the aforementioned identification, one needs to choose a function $t$ on the constraint surface for the role of time. Then, the constant time surfaces define the reduced phase space and the unconstrained formalism. This construction is depicted in Fig. \ref{int} (more details can be found in \cite{2M15}).  Therefore, general relativity relies intrinsically on an arbitrary choice of time $t$, which we shall call the internal clock. It is represented by a coordinate on the contact manifold and is subject to choice and transformations as much as the canonical coordinates. A very clear discussion of the meaning of the internal clock can be found in \cite{HK} (see especially the discussion on the Schwarzschild metrics).

Let us briefly mention the Dirac approach, which is a method for solving the constraint $C$, alternative to the reduced phase space approach. In this approach, one solves the quantum constraint operator equation,
\begin{align}\label{1con}
\hat{C}\Psi(q_1,\dots,q_n)=0.
\end{align}
For extracting dynamical content of the state $\Psi$, one ideally would like the operator $\hat{C}$ to be linear in some momentum, say $\hat{p}_1=i\frac{\partial}{\partial q_1}$,
$$\hat{C}=i\frac{\partial}{\partial q_1}-\hat{H},$$
where $\hat{H}$ does not involve differentiation with respect to $q_1$. Then, the $q_1$ becomes the internal clock. In the Page and Wootters proposal \cite{PaW,Lloyd} the ``flow" of time consists in the correlation between the quantum degree of freedom $\hat{q}_1$ and the rest of the system in the state $\Psi(q_1,\dots,q_n)$. A more common procedure would be to introduce the physical Hilbert space, $$\mathcal{H}_{phys}=L^2(\mathbb{R}^{n-1},\ud q_2\dots\ud q_n),$$ and reinterpret the constraint equation (\ref{1con}) as the Schr\"odinger equation that generates the dynamics in $\mathcal{H}_{phys}$ and with respect to the classical internal clock $q_1$. The solution to Eq. (\ref{1con}) is then accordingly reinterpreted, $$\Psi(q_1,\dots,q_n)=\psi(q_2,\dots,q_n)(t)\in\mathcal{H}_{phys},$$ for a fixed $t=q_1$. For clear reviews of the known approaches to quantization of Hamiltonian constraints, see \cite{Ku,Ish}.

Our approach is first to reduce the constraint $C$ by choosing an internal clock and then to quantize the respective reduced phase space. Therefore, we arrive directly at the Schr\"odinger equation instead of the quantum constraint equation (\ref{1con}). Provided the linear form of the constraint $C$ with respect to $p_1$, we may choose $q_1$ as the internal clock and define the reduced phase space as $$(q_2,\dots,q_n,p_2,\dots,p_n)\in\mathbb{R}^{2n-2}.$$ Then we can introduce the reduced Hamiltonian to this phase space, $$H=H(q_2,\dots,q_n,p_2,\dots,p_n),$$ which may also depend on the classical variable $t=q_1$; however, for simplicity, we assume that it does not. The reduced Hamiltonian generates the dynamics in the reduced phase space with respect to $q_1$. Its quantization leads to the Schr\"odinger equation in $\mathcal{H}_{phys}$, which in the simplest cases is equivalent to Eq. (\ref{1con}) as discussed above. In the following sections, we assume some initial choice of internal clock, the reduced phase space, and the reduced Hamiltonian and show how transformations of the internal clock can be implemented into the reduced phase space formalism and what their consequences are for quantum mechanics. Although, in the general case, the reduced phase space and the Dirac approach may yield different quantum dynamics (e.g., for more complex Hamiltonian constraints or for less obvious choices of the internal clock), they both involve the same arbitrariness in the internal clock. Therefore, we expect that the result of our work is, at least qualitatively, approach independent. \\ 

\begin{figure*}[t]
\centering
\includegraphics[width=0.9\textwidth]{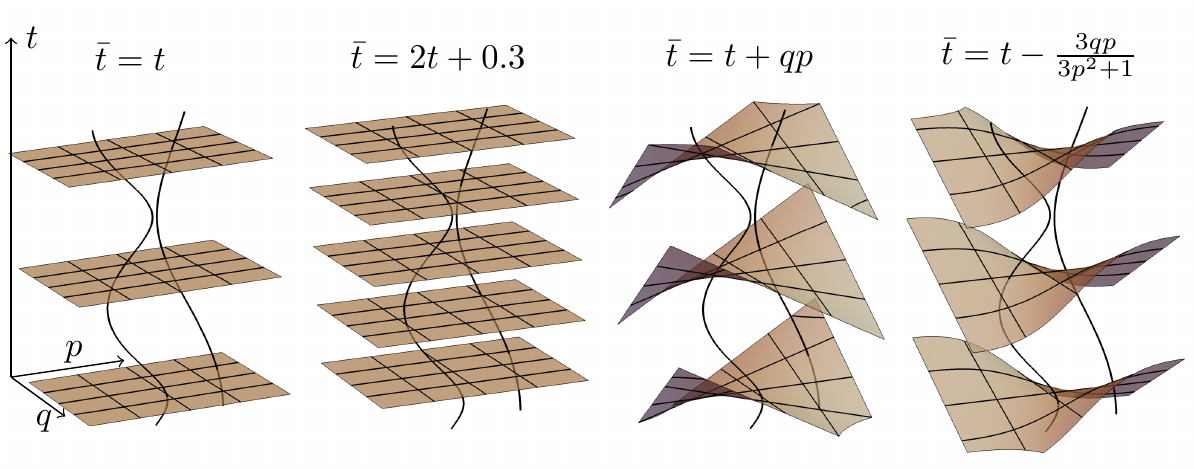}
\caption{The two curves (solid and dashed) represent the motion of a particle in a contact manifold (phase space $\times$ time manifold). The planes represent constant time surfaces which fix an abstract (from the point of view of classical mechanics) notion of simultaneity between states of a particle belonging to different solutions (represented here by curves). The second picture from the left illustrates a time transformation which merely changes time units and the zero-time point. The third and fourth pictures from the left illustrate time transformations which change simultaneity between states of a particle belonging to different curves; they are studied in detail in Sec. \ref{eg2}. }
\label{2}
\end{figure*}

Let us recall the canonical description of the nonrelativistic mechanics based on an external and fixed variable $t$ which parametrizes the evolution of systems. For each value of the time variable $t$ there corresponds a state of a system in the phase space. Hence, the motion is represented by parametrized curves that lie in the phase space extended by the time dimension $(q,p,t)$. The absolute time variable fixes a foliation of the $(q,p,t)$ manifold. Every slice of constant time admits a symmetry given by canonical transformations that preserve the Poisson brackets between observables. 

Taking seriously the idea of internal clocks, we shall put $t$ on the same footing as the remaining internal variables, the canonical variables. The way to do so is to extend the symmetry given by canonical transformations to transformations that do not preserve the internal clock $t$. In other words, we search for a new formalism that is invariant under transformations of internal clocks, which in general take the form
\begin{align}\label{ict}
t \mapsto \bar{t}=\bar{t}(t,q,p).
\end{align}
See Fig. \ref{2} for some examples. We postpone the task of formulating such a formalism to later sections. At present let us notice that such transformations are not canonical because they do not preserve constant time surfaces (see Fig. \ref{2}) as the canonical transformations do. In this light, we have grounds to expect that the canonical structure, which is usually given by the Poisson bracket, is different for different internal clocks. This does not make a difference at the classical level where the Poisson bracket is an auxiliary, nonobservable structure. On the other hand, the role of the Poisson bracket is completely basic for constructing a quantum theory. Therefore, we anticipate some nontrivial effects related to the change of clock with respect to which we measure the evolution, namely, the clock effects (for a recent study of the clock effect, see \cite{MB}). How quantum theories based on different internal clocks are related and how to put them into a common framework is the subject of the present paper.\\

Broadly speaking, the present work is a proposal for a reformulation of quantum mechanics in such a way as to replace the fixed and external time by arbitrary and internal clocks. We place all the quantum descriptions involving all the possible clocks in a fixed Hilbert space. Their basic property is the invariance of nondynamical properties of quantum systems with respect to transformations of the internal clock. Also, the quantum evolution of physical states is shown to be unique for all clocks. The only difference with respect to ordinary quantum mechanics turns out to be the interpretation of dynamical properties of physical states, which depends on the choice of internal clock. As we show in the end of the article, our proposal contains quantum mechanics in its usual formulation as a special case. This, to our mind, provides a strong argument in favor of our formalism. \\ 

The outline of the paper is as follows. Section \ref{II} recalls the notion of canonical transformations and describes an extension called pseudocanonical transformations, which include the internal clock transformations (\ref{ict}). Section \ref{eg1} illustrates the extended theory with the canonical formalism of a particle on the real line. Section \ref{IV} describes the effect of pseudocanonical transformations on the form of quantum theory. The discussion is based on the assumption of a unique quantization map underlying the quantum theory in all internal clocks and leading to a unique representation of constants of motion. The results of this discussion are illustrated again with a particle on the real line in Section \ref{eg2}. In Section \ref{VI} we describe how the usual form of quantum mechanics can be obtained from our formulation. We conclude in Section \ref{VII}. The Appendix provides some technical details of the analysis made in Section \ref{eg2}.

\section{Canonical formalism and clocks}\label{II}
Let us recall the basic framework of classical mechanics. We consider a phase space $(q,p) \in \mathbb{R}^2$ equipped with the symplectic form $\omega=\ud q\ud p$ and the Hamiltonian $H(q,p)$, which generates dynamics in $t$ through the Hamilton equations:
\begin{equation}\label{oldHE}
\frac{\ud q}{\ud t}=-\omega^{-1}(q,H),~~~ \frac{\ud p}{\ud t}=-\omega^{-1}(p,H),
\end{equation}
where the Poisson bracket is the (minus) inverse of the symplectic form, $\{\cdot,\cdot\}=-\omega^{-1}(\cdot,\cdot)$. This framework is conveniently reformulated by means of the contact manifold which is the product of the phase space and the clock manifold,
\begin{align}(q,p,t)\in\mathbb{R}^3.\end{align}
The symplectic form is then replaced by the contact form,
\begin{align}\omega_{C}=\ud q\ud p-\ud t\ud H(q,p),\end{align}
which lives in the contact manifold $\mathbb{R}^3$ rather than the phase space $\mathbb{R}^2$. It reduces to the symplectic form at constant clock surfaces,
\begin{align}\label{symcon}
\omega_{C}|_t=\omega=\ud q\ud p.\end{align}
Therefore, the definition of the Poisson bracket is modified accordingly as the (minus) inverse of the contact form for constant clock surfaces, 
\begin{align}\{\cdot,\cdot\}=-\omega_C|_t^{-1}(\cdot,\cdot).\end{align}

\subsection{Canonical transformations}
Now let us recall the very useful concept of canonical transformations (see, e.g., \cite{AbMa}). They are defined as passive transformations of the contact manifold,
\begin{align}\mathbb{R}^3\ni (q,p,t)\mapsto (\bar{q},\bar{p},t)\in \mathbb{R}^3,\end{align}
such that the contact form keeps its original form, that is,
\begin{align}\omega_C=\ud\bar{q}\ud\bar{p}-\ud t\ud \bar{H} (\bar{q},\bar{p}).\end{align}
Note that the internal clock $t$ is preserved by those transformations. This makes sense if the Poisson bracket is to be preserved. Note that the constant clock surfaces $t$ and the contact form $\omega_C$ are unique geometrical objects and so is $\omega_C|_t$. Thus, the symplectic forms,
\begin{align}\ud q\ud p=\ud\bar{q}\ud\bar{p},\end{align}
are equal, and their inverses define a unique Poisson bracket. These transformations constitute a symmetry of the classical formalism. Since the canonical formalism is the one used for quantization it is not surprising that this symmetry exists to some extent at the quantum level in the form of unitary transformations. 

\subsection{Pseudocanonical transformations}

The idea of pseudocanonical transformations is to incorporate the freedom of choosing the internal clock in the canonical formalism as much as canonical transformations incorporate the freedom in choosing the
canonical coordinates \cite{M}.

Given initial canonical coordinates and the internal clock $(q,p,t)$, in terms of which the contact form reads
\begin{align}\label{oldCF}
\omega_{C}=\ud q\ud p-\ud t\ud H,
\end{align}
pseudocanonical transformations are defined as such transformations of the canonical variables and the internal clock,
\begin{align}\mathbb{R}^3\ni (q,p,t)\mapsto (\bar{q},\bar{p},\bar{t})\in \mathbb{R}^3,\end{align}
that the form of the contact form is preserved,
\begin{align}\label{newCF}\omega_C=\ud\bar{q}\ud\bar{p}-\ud \bar{t}\ud \bar{H} (\bar{q},\bar{p}).\end{align}
Observe that contrary to the case of canonical transformations, the internal clock is freely transformed. The only extra requirement is that the clock $\bar{t}$ is monotonic with respect to $t$ and can be written as:
\begin{align}\frac{\ud\bar{t}}{\ud t}=\frac{\partial\bar{t}}{\partial t}-\omega^{-1}(\bar{t},H)>0,\end{align}
where $-\omega^{-1}(\cdot,\cdot)$ is the Poisson bracket associated with the clock $t$. Now, notice that 
\begin{align}\omega_C|_{\bar{t}}\neq\omega_C|_t,\end{align}
as $\bar{t}$ and $t$ define different constant clock submanifolds in the contact manifold. So we have a new symplectic form
\begin{align}
\bar{\omega}=\omega_C|_{\bar{t}}.\end{align}
The equations of motion again take the form of the Hamilton equations,
\begin{equation}\label{newHE}
\frac{\ud \bar{q}}{\ud \bar{t}}=-\bar{\omega}^{-1}(\bar{q},\bar{H}),~~~ \frac{\ud \bar{p}}{\ud \bar{t}}=-\bar{\omega}^{-1}(\bar{p},\bar{H}),
\end{equation}
but with respect to the new clock $\bar{t}$.

\subsection{Special pseudocanonical transformations}

The defining requirement that the pseudocanonical transformations preserve the form of the contact form as shown in Eq. (\ref{newCF}) can be further specialized. Namely, we may demand that the new Hamiltonian,
\begin{align}\bar{H}(\bar{q},\bar{p})=H(\bar{q},\bar{p}),\end{align}
exhibits the same formal dependence on the basic variables as the initial one, $H(q,p)$. Since the basic variables are canonical with respect to their respective Poisson brackets,
\begin{align}-\omega^{-1}(q,p)=1=-\bar{\omega}^{-1}(\bar{q},\bar{p}),\end{align}
it follows that the Hamilton equations (\ref{oldHE}) and (\ref{newHE}) must now be formally the same. Therefore, given a solution to the initial Hamilton equations,
\begin{align}q=Q(t),~~p=P(t),\end{align}
the respective solution to the new Hamilton equations exists,
\begin{align}\bar{q}=Q(\bar{t}),~~\bar{p}=P(\bar{t}),\end{align}
where we have simply replaced $t$ with $\bar{t}$ in the formal expression for the solution to obtain the same solution parametrized by $\bar{q}$, $\bar{p}$, and $\bar{t}$. 

It follows that the observable
\begin{align}C_J(q,p,t)\end{align}
is a constant of motion, i.e. $\partial_tC_J=\omega^{-1}(C_J,H(q,p))$, if and only if 
\begin{align}C_J(\bar{q},\bar{p},\bar{t})\end{align}
is a constant of motion too, i.e. $\partial_{\bar{t}}C_J=\bar{\omega}^{-1}(C_J,H(\bar{q},\bar{p}))$. Therefore, we will specialize pseudocanonical transformations to the so-called special pseudocanonical transformations that preserve the form of all constants of motion,
\begin{align}\label{SPT1}
C_J(q,p,t)=C_J(\bar{q},\bar{p},\bar{t}).
\end{align}
The above relation is a set of $2n$ algebraic equations as there are exactly that many independent constants of (integrable) motion in $2n$-dimensional phase space. Equation (\ref{SPT1}) supplemented with the clock transformation,
\begin{align}\label{SPT2}
\bar{t}=\bar{t}(q,p,t),
\end{align}
fixes a new $2n+1$-dimensional system of contact coordinates $(\bar{q},\bar{p},\bar{t})$.

Clock transformations can be most generally expressed in terms of the delay function,
\begin{align}t\mapsto\bar{t}=t+D(q,p,t).\end{align}
In what follows, for the sake of simplicity, we will assume that the delay function is slightly restricted, namely\begin{align}
D=D(q,p).
\end{align}

\section{Example: free particle I} \label{eg1}
In what follows we illustrate the above formalism with an example of a free particle on the real line:
\begin{equation}\label{defP}
\omega_C=\ud q\ud p-\ud t\ud H,~~H=\frac{p^2}{2},
\end{equation}
where $(q,p)\in \mathbb{R}^2$ and $t \in \mathbb{R}$.\\

To introduce special pseudocanonical transformation, first, we identify all the independent constants of motion. They are
\begin{equation}
C_1(q,p,t)=p,~~~C_2(q,p,t)=q-pt
\end{equation}
The special pseudocanonical transformation is given by the set of algebraic relations
\begin{equation}
\begin{split}
C_1(q,p,t)&=C_1(\bar{q},\bar{p},\bar{t}),\\
C_2(q,p,t)&=C_2(\bar{q},\bar{p},\bar{t}),\\
\bar{t}&=t+D(q,p),
\end{split}
\end{equation}
which are equivalent to
\begin{equation}\label{psP}
\begin{split}
\bar{p}&=p,\\
\bar{q}&=q-p D(q,p),\\
\bar{t}&=t+D(q,p).
\end{split}
\end{equation}
By explicit substitution one can verify that the form of the contact form (\ref{defP}) is indeed preserved by (\ref{psP})
\begin{equation}
\omega_C=d\bar{q}d\bar{p}-d\bar{t}d\bar{H},~~\bar{H}=\frac{\bar{p}^2}{2}.
\end{equation}
Although the contact form remains the same before and after the transformation, it is easily verified that the symplectic form,
\begin{align}\omega:=\omega_C|_t\neq\omega_C|_{\bar{t}}=:\bar{\omega},\end{align}
is modified and
\begin{equation}
-\omega^{-1}(\bar{q},\bar{p})=-\omega^{-1}\left(q-p D(q,p),p\right)
\neq-\bar{\omega}^{-1}(\bar{q},\bar{p})=1.
\end{equation}
The new clock must be monotonic with respect to the old one,
\begin{equation}\label{mono}
\frac{\ud\bar{t}}{\ud t}=1+p\frac{\partial D}{\partial q}>0.
\end{equation}
The solution to the above equation defines a family of admissible delay functions which neither stop nor invert the flow of evolution.
\section{Quantum mechanics and clocks}\label{IV}

Let us assume quantization as a linear map from functions on phase space to linear operators in the Hilbert space $\mathcal{H}$, which in the most general case reads \cite{JPGconf1,JPGconf2}
\begin{align}
f(q,p,t)\mapsto A_f:=\int_{t=const} \ud q\ud p~ f(q,p,t) M(q,p),
\end{align}
where the integration is made over the particular phase space determined by the level sets of a given clock and $M(q,p)$ is a family of bounded operators on $\mathcal{H}$, which resolves the identity,
\begin{align}
\int_{t=const} \ud q\ud p~ M(q,p)=\id.
\end{align}
This broad definition includes the case of the Weyl-Wigner quantization (i.e., the canonical prescription), 
\begin{equation}\label{WW}
M(q,p)=\mathbf{D}(q,p)2\mathcal{P}\mathbf{D}^{\dagger}(q,p),
\end{equation}
where $\mathcal{P}$ is the parity operator, $\mathbf{D}(q,p)=e^{i(p\hat{Q}-q\hat{P})}$ is the displacement operator, and $\hat{Q}$ and $\hat{P}$ are the position and momentum operators \cite{WHcomp}. Suppose that we work in another canonical formalism given in terms of $(\bar{q},\bar{p},\bar{t})$, in which we define quantization analogously,
\begin{align}
f(\bar{q},\bar{p},\bar{t})\mapsto \bar{A}_f:=\int_{\bar{t}=const} \ud \bar{q}\ud \bar{p}~ f(\bar{q},\bar{p},\bar{t}) \bar{M}(\bar{q},\bar{p}),
\end{align}
where $\bar{M}(\bar{q},\bar{p})$ resolves the identity, too. Note that in this case the integration is taken over different phase spaces determined by a new clock, $\bar{t}=const$. There is no relation between $\bar{M}(\bar{q},\bar{p})$ and $M(q,p)$ at the moment. The question that we address in the present article is the question of dissimilarities between quantum theories obtained through some acceptable quantization maps defined in canonical formalisms based on different internal clocks. We are interested in the dissimilarities due to different choices of the internal clock rather than usual quantization ambiguities that may arise even in a fixed time formalism like ordering ambiguity. The principal question is how to ensure that the latter does not obscure the former.

There exists a natural way to handle the above issue. Namely, one requires that all the observables that are constants of motions are quantized in the same way irrespectively of the internal clock in which the quantization map is defined. This makes sense because one expects that any nondynamical quantum property of a given system should not depend on the clock employed solely for describing its evolution. We notice that the special pseudocanonical transformations introduced by Eq. (\ref{SPT1}) are such that observables that are constants of motion have exactly the same dependence on the internal clock and the respective canonical coordinates. Thus, we require that
$$\int_{t=const} \ud q\ud p~ C_J(q,p,t) M(q,p)$$
$$=\int_{\bar{t}=const} \ud \bar{q}\ud \bar{p}~ C_J(\bar{q},\bar{p},\bar{t}) \bar{M}(\bar{q},\bar{p}),$$
where $C_J$ is any nondynamical observable and the equality holds for any $t=\bar{t}$ numerically (though the integration is made over different phase spaces). Since there are exactly $2n$ independent constants of motions, which is equal to the dimensionality of the phase space, we conclude that
\begin{equation}\label{MM}
M(\cdot,\cdot)\equiv\bar{M}(\cdot,\cdot).\end{equation}
It follows that observables that have formally the same dependence on the respective contact coordinates are promoted to unique operators, or unique families of operators enumerated by the values of the respective clocks, namely,
\begin{align}f(q,p)\mapsto \hat{F}~~\Rightarrow ~~f(\bar{q},\bar{p})\mapsto \hat{F}\end{align}
or
\begin{align}f(q,p,t)\mapsto \hat{F}_t~~\Rightarrow ~~f(\bar{q},\bar{p},\bar{t})\mapsto \hat{F}_{\bar{t}}~.\end{align}

Now we shall consider two cases. The first case is when the observable $f(q,p,t)$ is a constant of motion. Then it is promoted to a unique quantum operator, say $\hat{C}$. Provided that it is self-adjoint, there exists a unique wave function corresponding to any state $|\psi\rangle\in\mathcal{H}$,
\begin{align}\label{psiC}
\langle \phi_c|\psi\rangle=\psi(c)\in L^2(sp(\hat{C}),\ud c),~~ \hat{C}|\phi_c\rangle=c|\phi_c\rangle,
\end{align}
where $c$ denotes the eigenvalues of $\hat{C}$. Since the operator $ \hat{C}$ has a unique physical interpretation in all internal clocks, the unique wave function of Eq. (\ref{psiC}) provides a unique physical interpretation to the state $|\psi\rangle\in\mathcal{H}$.

The second case is when the observable $f(q,p,t)$ is not a constant of motion. Then, for another choice of clock, say $\bar{t}$, we find that
\begin{align}f(q,p,t)=g(\bar{q},\bar{p},\bar{t}),~\textrm{where}~f(\cdot,\cdot,\cdot)\neq g(\cdot,\cdot,\cdot),\end{align}
under the transformation set by Eqs. (\ref{SPT1}) and (\ref{SPT2}). Upon quantization, they will be assigned two distinct operators, say $\hat{F}_t$ and $\hat{G}_{\bar{t}}$, respectively. Provided that they are self-adjoint, there exist two distinct wave functions corresponding to any state $|\psi\rangle\in\mathcal{H}$,
\begin{align}
\langle \phi_f|\psi\rangle=\psi(f)\in L^2(sp(\hat{F}_t),\ud f),~~ \hat{F}_t|\phi_f\rangle=f|\phi_f\rangle,
\end{align}
and
\begin{align}
\langle \bar{\phi}_g|\psi\rangle=\bar{\psi}(g)\in L^2(sp(\hat{G}_{\bar{t}}),\ud g),~~ \hat{G}_{\bar{t}}|\bar{\phi}_g\rangle=g|\bar{\phi}_g\rangle,
\end{align}
where $f$ and $g$ denote the eigenvalues of $\hat{F}_t$ and $\hat{G}_{\bar{t}}$, respectively. However, $f$ and $g$ have exactly the same physical meaning in their respective clock-based formalisms, and since the spectra might not be the same, or at least,
\begin{align}\psi(\cdot)\neq\bar{\psi}(\cdot),\end{align}
we conclude that a single state $|\psi\rangle\in\mathcal{H}$ is given distinct dynamical interpretations in those two formalisms.

Furthermore, we notice that the Hamiltonian, $H(q,p)$ or $H(\bar{q},\bar{p})$, is formally the same function in all clocks and thus, it is promoted to a unique quantum Hamilton operator, say $\hat{H}$. Therefore, there is a unique Schr\"odinger equation governing the evolution of quantum states,
\begin{equation}\label{sch}
i\hbar \frac{\partial}{\partial \tau} |\psi \rangle=\hat{H}| \psi \rangle,
\end{equation}
where $\tau\equiv t$ or $\tau\equiv\bar{t}$. Hence, up to parametrization, the solution to Eq. (\ref{sch}) is unique. 

To conclude this section, we summarize the  interesting properties exhibited by the proposed framework:
\begin{description}
\item[i] ~~Given a physical system, all the respective canonical formalisms based on all possible internal clocks and related by pseudocanonical transformations may be quantized in a uniform manner. All the respective quantum theories may be placed in the same Hilbert space $\mathcal{H}$. 
\item[ii] ~~Any nondynamical information about any state $|\psi\rangle\in\mathcal{H}$  is provided by means of spectral decomposition induced by a nondynamical operator and is completely independent of the choice of internal clock.
\item[iii] ~~Unitary evolution of any initial state 
\begin{align}\mathbb{R}\ni\tau\mapsto |\psi(\tau)\rangle\in\mathcal{H}\end{align} is completely independent of the choice of internal clock. 
\item[iv] ~~Any dynamical information about any state $|\psi\rangle\in\mathcal{H}$ is provided by means of spectral decomposition induced by a dynamical operator and depends crucially on the choice of internal clock.
\item[v] ~~Interpretation of the evolution of any initial state in terms of spectral decomposition induced by a self-adjoint dynamical $\hat{F}$,
\begin{align}\mathbb{R}\ni\tau\mapsto \langle\phi_f|\psi(\tau)\rangle=\psi(f,\tau)\in L^2(sp(\hat{F}),\ud f),\end{align} 
 depends crucially on the choice of internal clock.
\end{description}

\section{Example: free particle II} \label{eg2}
In what follows we extend the classical-level analysis of a free particle, which we made in Sec. \ref{eg1} to the quantum level by making use of the framework presented in the last section. First, we use the canonical quantization prescription to obtain the position and the momentum operators. The momentum operator represents a conserved quantity, and thus we use it to give a choice of clock-independent characterization of state vectors in the Hilbert space. The position operator represents a dynamical quantity and thus, we use it to study the dependence of the respective characterisations of state vectors on the choice of internal clock. 

Suppose that $q$ represents the position of a particle, $p$ represents the momentum of a particle, and $t$ represents some clock. Now, $\bar{q}$, $\bar{p}$, and $\bar{t}$ are another set of contact coordinates that are related to the original one by the pseudocanonical transformation (\ref{psP}) after exchanging the roles of the barred and unbarred variables. We quantize the canonical coordinates by the Weyl-Wigner map (\ref{WW}), which is applied to all possible phase spaces by the formal replacement of the respective canonical coordinates in accordance with Eq. (\ref{MM}). For the clock $t$, we obtain
\begin{equation}\label{quantization1}
\begin{split}
q&\mapsto\hat{Q},\\
p&\mapsto\hat{P}, 
\end{split}
\end{equation}
and for the clock $\bar{t}$,
\begin{equation}\label{quantization2}
\begin{split}
q=\bar{q}-\bar{p}D(\bar{q},\bar{p})&\mapsto Sym[\hat{Q}-\hat{P}D(\hat{Q},\hat{P})],\\
p=\bar{p}&\mapsto\hat{P}, 
\end{split}
\end{equation}
where $\hat{Q}$ and $\hat{P}$ are, respectively, the position and the momentum operator on the real line and $Sym[\cdot]$ denotes symmetrisation of the expression $[\cdot]$ with respect to the basic operators. We observe in consistency with the discussion of the last section that the conserved momentum $p$ is promoted the same operator irrespectively of the choice of internal clock, whereas the dynamical position $q$ is promoted to different operators in different clocks.

Since the momentum $p$ is promoted to the same operator $\hat{P}$ in all clocks, we can ascribe a unique wave function to any state vector $|\psi\rangle\in\mathcal{H}$,
\begin{equation}
\hat{\psi}(p)=\langle p|\psi\rangle,
\end{equation}
where $\hat{P}|p\rangle=p|p\rangle$. The associated probability distribution,
\begin{align}P_{\hat{\psi}}(p)=|\psi(p)|^2,\end{align}
provides a unique distribution of momenta in a given state vector $|\psi\rangle\in\mathcal{H}$ irrespectively of the choice of clock with respect to which we measure the momentum. Therefore, we will use the momentum operator to fix a unique state by means of a wave function in $p$ and then study its spectral decomposition with respect to dynamical operators. We set the wave function
\begin{equation}\label{state}
\hat{\psi}(p)=(\pi \sigma)^{-1/4}e^{-\frac{1}{2\sigma} (p-p_0)^2}e^{-i x_0 p},
\end{equation}
where $x_0$ is the expectation value of the position operator (in clock $t$), $p_0$ is the expectation value of the momentum operator, and $\sigma$ is the dispersion of momentum. 

Now, our task is to represent the state (\ref{state}) as a wave function on the spectrum of the position operator corresponding to the classical observable $q$. Since we have different operators corresponding to this observable depending on the choice of clock, we obtain many different respective eigenvalue problems. In the clock $t$, it reads
\begin{equation}\label{qt}
\hat{Q}|q\rangle=q|q\rangle ,
\end{equation}
whereas in the clock $\bar{t}$, it reads
\begin{equation}\label{qbart}
Sym[\hat{Q}-\hat{P}D(\hat{Q},\hat{P})]|q\rangle=q|q\rangle .
\end{equation}
Since the solutions $|q\rangle$ to Eqs. (\ref{qt}) and (\ref{qbart}) are different, the associated wave functions,
\begin{equation}
\psi(q)=\langle q|\psi\rangle,
\end{equation}
will be different, too. In fact, even the spectrum can be different in character, continuous or discrete, depending on the choice of clock. Let us denote the spectrum as $\mathbb{X}=\{q\}$. Let us also constrain the delay functions $D$ to the following form:
\begin{align}\label{delayP}D(\bar{q},\bar{p})=\bar{q}f(\bar{p}).\end{align}
Then, it can be deduced from Eq. (\ref{quantization2}) that the position operator in the clock $\bar{t}$ acts on wave functions in momentum representation $\hat{\psi}(p)\in L^2(\mathbb{R},\ud p)$ as follows: 
\begin{equation} \label{new_operators}
q \mapsto  i \frac{\partial}{\partial p}+\frac{1}{2}\left( i \frac{\partial}{\partial p} \cdot pf(p)+pf(p) \cdot i \frac{\partial}{\partial p} \right).
\end{equation}
In the Appendix, we solve the eigenvalue problem for the above operator explicitly and show that there are two possible cases:
$$ \mathbf{A.} \ \int_{-\infty}^{+\infty} \frac{\ud p}{1+pf(p)} < \infty $$

$$ \mathbf{B.} \ \int_{-\infty}^{+\infty} \frac{\ud p}{1+pf(p)} = \infty.$$
In case \textbf{A.} the spectrum $\mathbb{X}$ is discrete, whereas in case \textbf{B.} the spectrum $\mathbb{X}$ is continuous.

\begin{figure*}[t]
\centering
\includegraphics[width=0.45\textwidth]{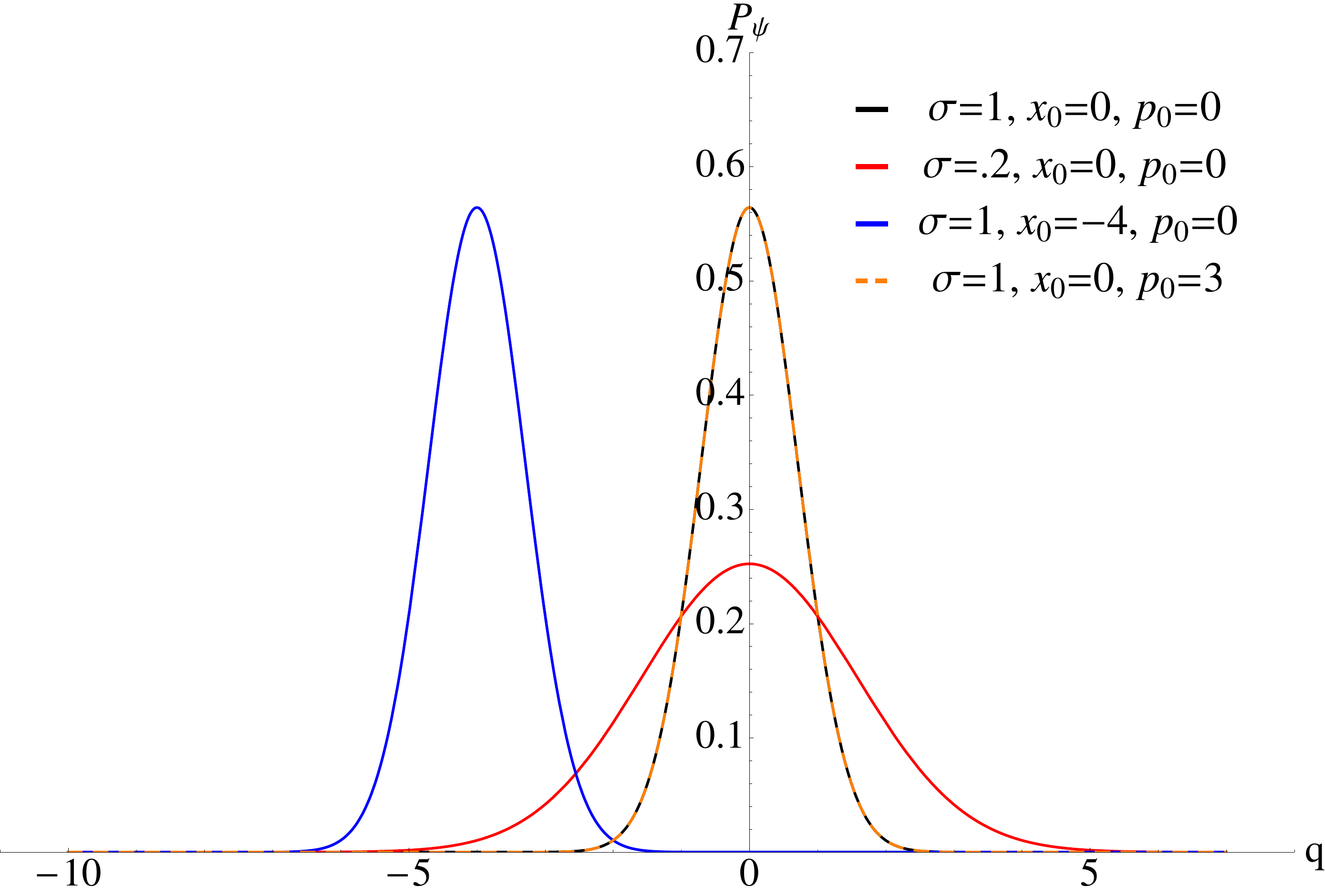}
\includegraphics[width=0.45\textwidth]{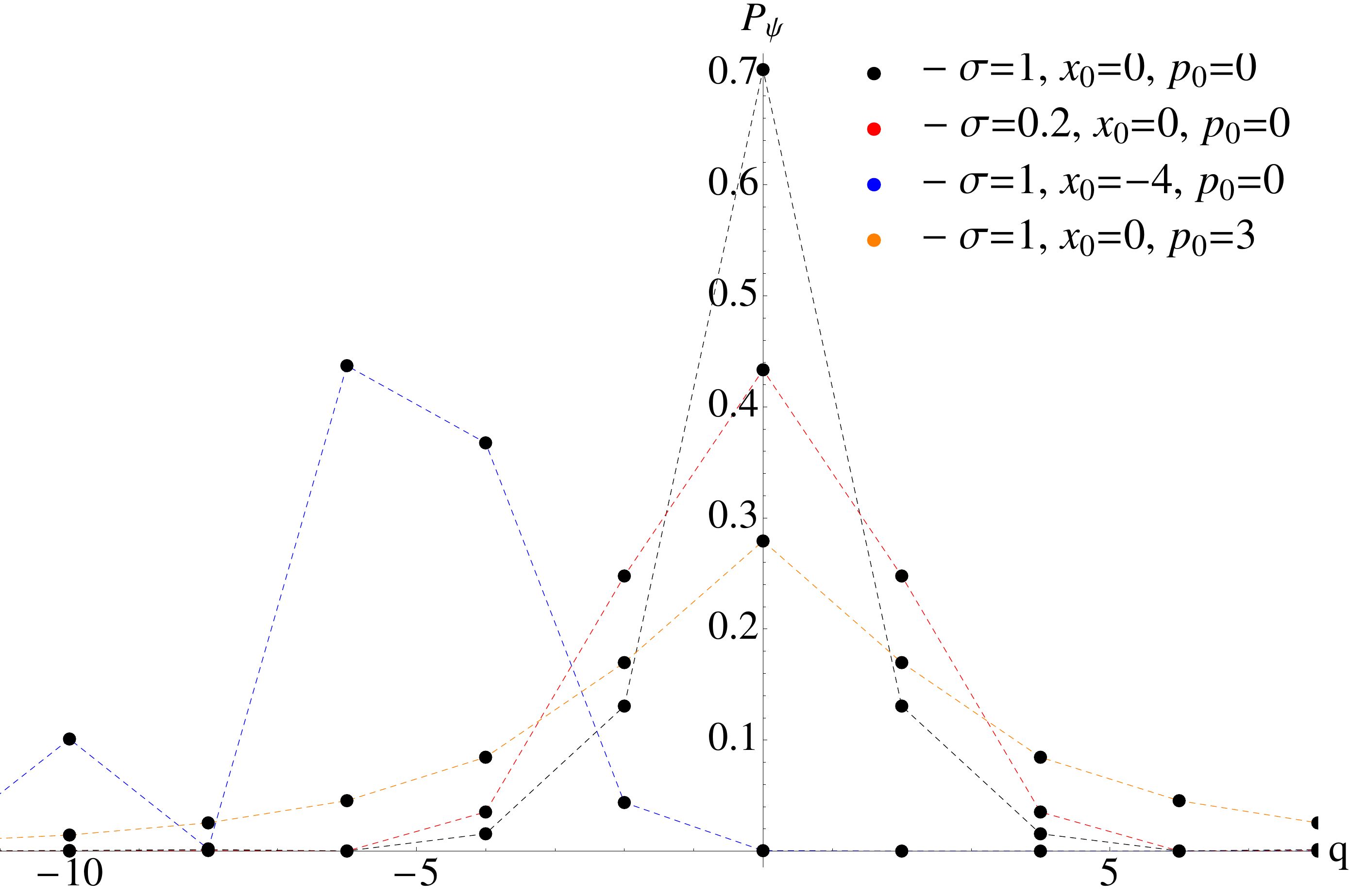}
\caption{Probability distribution $P_{\psi}=|\langle q | \psi \rangle|^2$ of position eigenvalues for the state $| \psi \rangle$ [defined in (\ref{state})] in the old clock $t$ (on the left) and in the new clock $\bar{t}=t+qp$ (on the right). On the right: the probability for specific eigenvalues is marked with dots. The spectrum is discrete.}
\label{discrete_image}
\end{figure*}

\subsection{Discrete spectrum}
We consider an example of a delay function (see Fig. \ref{2}),
\begin{equation}
D(\bar{q},\bar{p})=\bar{q}\bar{p}.
\end{equation}
The monotonicity condition (\ref{mono}) is trivially fulfilled, and we find from Eq. (\ref{new_operators})
\begin{equation}
\begin{split}
q &\mapsto i(1+p^2) \frac{\partial}{\partial p}+ip,\\
|q\rangle &=\frac{e^{-iq\arctan(p)}}{\sqrt{\pi}\sqrt{p^2+1}},~~q =2 n + \mu\in\mathbb{X},
\end{split}
\end{equation}
where $n\in \mathbf{N}$ enumerates the eigenvalues and $\mu \in [0,2)$ is a fixed parameter that enumerates unitarily inequivalent representations and we choose $\mu=0.$\\
It immediately follows that the wave function in $q$ corresponding to the state vector $|\psi\rangle$ of Eq. (\ref{state}) reads
\begin{equation}
\begin{split}
\psi(q)=\langle q | \psi \rangle=\int_{\mathbb{R}}\frac{\ud p}{\sqrt{p^2+1}}\frac{e^{iq\arctan(p)}}{\sqrt{\pi}}~\hat{\psi}(p).
\end{split}
\end{equation}
We plot the associated probability distribution $P_{\psi}=|\langle q | \psi \rangle|^2$ in Fig. \ref{discrete_image}. The spectrum $\mathbb{X}$ is discrete, the probability of measuring a specific position is marked with a dot, the straight dashed lines connecting the dots do not have any physical meaning, and they simply interpolate the distribution domain to all real values. 

It can be seen that the states of vanishing average position $x_0=0$ in the original clock $t$ correspond to distributions with zero average position in the clock $\bar{t}$. Moreover, the dispersion of position distribution also increases with lowering the dispersion of momentum distribution $\sigma$. However, the specific dependence is now different. In particular, contrary to the original case, for the clock $\bar{t}$, the larger the value of $p_0$ (the average momentum) the larger the dispersion of position for a fixed dispersion of momentum. 

The case of nonvanishing average position $x_0=-4$ is even more interesting. One can clearly see that the average position remains nonzero; nevertheless, its value is now different as $\langle q \rangle \approx -6 \neq x_0$. The dispersion of position is also modified as $\sum_{q} P_{\psi} \cdot \sqrt{(q-\langle q \rangle)^2}\approx 1.82 \neq 1$. What is even more striking, the probability distribution possesses more than one maximum contrary to the original case.
 
\subsection{Continuous spectrum}
This case is divided into two subcases: 

\textit{1.} One of the integrals $\ \int_{0}^{\pm\infty} \frac{\ud p}{1+pf(p)} $ is finite.

\textit{2.} Neither of the above integrals is finite. 

The latter corresponds to the position operator with the real spectrum, whereas the former corresponds to the position operator, which is not self-adjoint. Thus, we study subcase \textit{2.}\\

\begin{figure*}[t]
\centering
\includegraphics[width=0.45\textwidth]{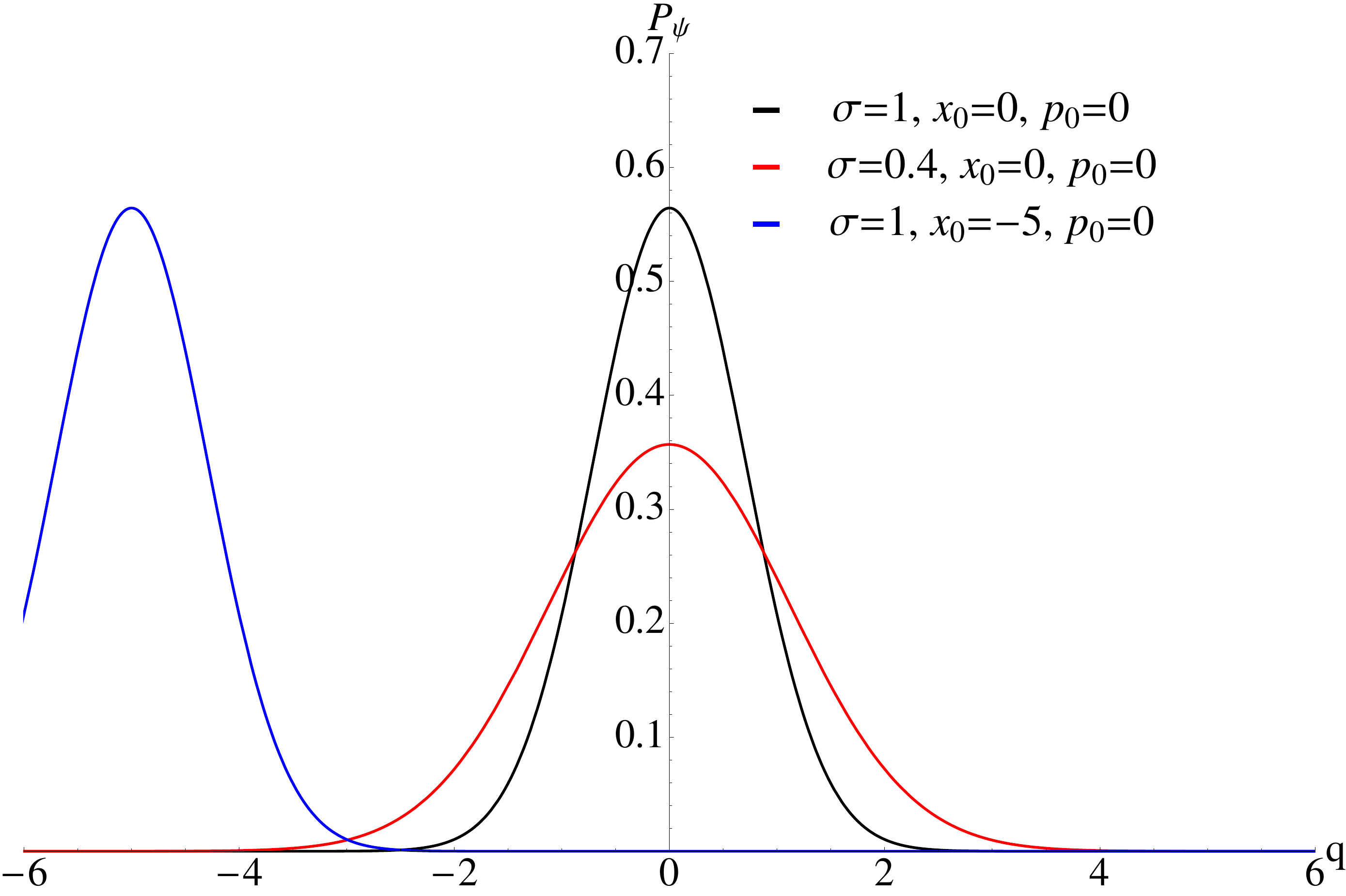}
\includegraphics[width=0.45\textwidth]{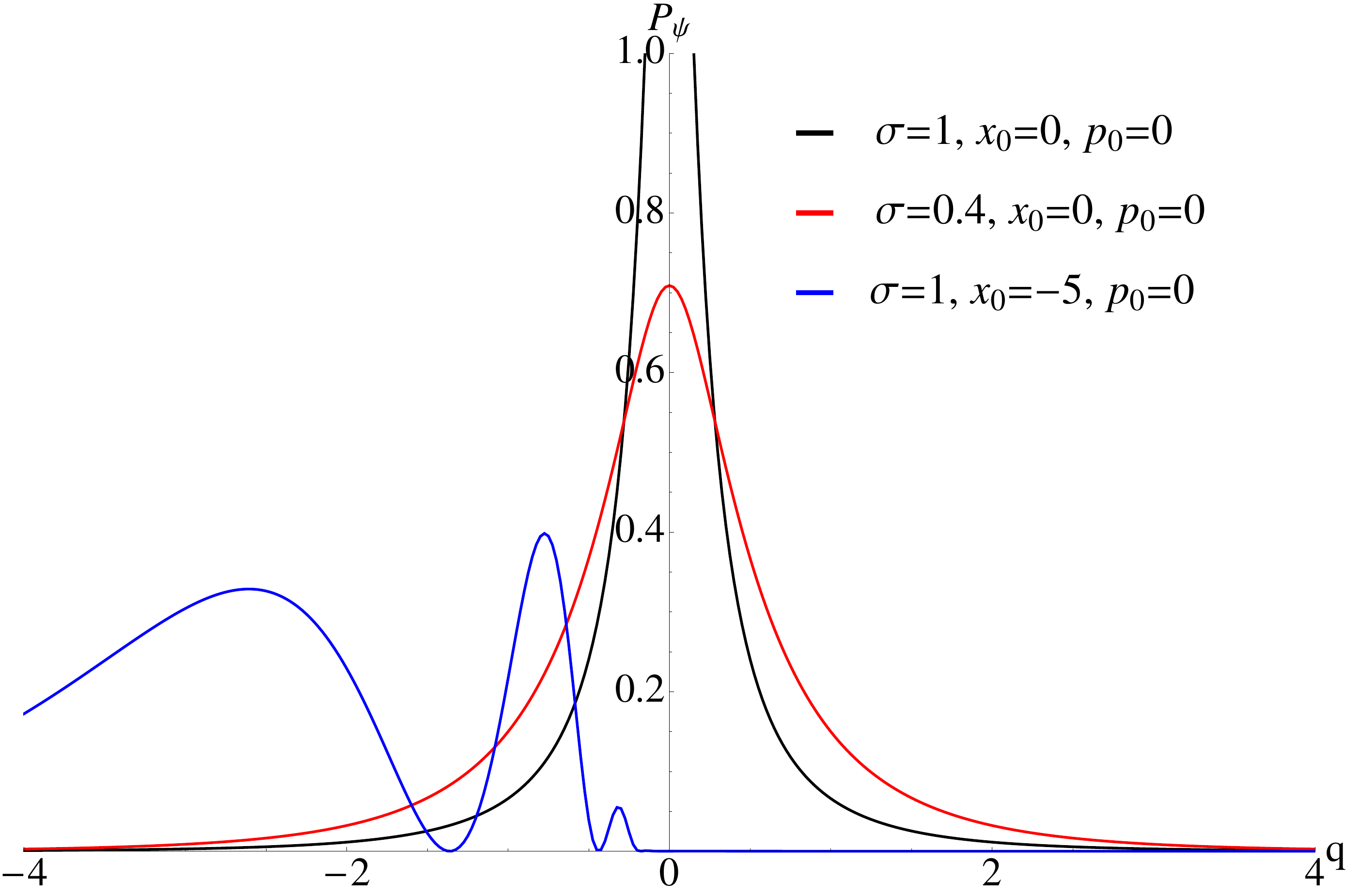}
\caption{Probability distribution $P_{\psi}=|\langle q | \psi \rangle|^2$ of position eigenvalues for the state $| \psi \rangle$ [defined in (\ref{state})] in the old clock $t$ (on the left) and in the new clock $\bar{t}=t-3\frac{qp}{1+3p^2}$ (on the right). The spectrum is continuous in both cases.}
\label{spectrumc}
\end{figure*}

We consider the delay function (see Fig. \ref{2})
\begin{equation}
D(\bar{q},\bar{p})=-\frac{3\bar{q}\bar{p}}{3 \bar{p}^2+1}
\end{equation}
The monotonicity condition (\ref{mono}) is trivially fulfilled, and according to Eq. (\ref{new_operators}), the position operator in the new clock and its eigenvectors read
\begin{equation}
\begin{split}
q &\mapsto \frac{i}{3p^2+1} \frac{\partial}{\partial p}-i\frac{3p}{(3p^2+1)^2},\\
|q\rangle&=\sqrt{\frac{3p^2+1}{2 \pi}}e^{-iq(p^3+p)},~~q\in\mathbb{R}.
\end{split}
\end{equation}
The spectrum of the position operator in the new clock is real $\mathbb{X}=\mathbb{R}$.

It immediately follows that the wave function in $q$ corresponding to the state vector $|\psi\rangle$ of Eq. (\ref{state}) reads
\begin{equation}
\begin{split}
\psi(q)=\langle q | \psi \rangle=\int_{\mathbb{R}} \ud p \sqrt{3p^2+1} e^{iq(p^3+ap)}~\hat{\psi}(p).
\end{split}
\end{equation}
It can be seen from Fig. \ref{spectrumc} that for all the included values of the parameters of the state vector ($\sigma$, $x_0$, $p_0$) the respective probability distributions $P_{\psi}=|\langle q | \psi \rangle|^2$ in the new and the old clocks can differ significantly. For $x_0=0$, the dispersion of position in the new clock is visibly smaller, though the vanishing average position is preserved. For the nonvanishing $x_0=-5$, the character of the distribution has changed a lot. In particular, it exhibits more than one maximum.

Some previous results on physical operators in the Dirac approach to quantization show that it is impossible to predict the character of their spectra from their definitions in the kinematical Hilbert space which does not yet satisfy the quantum constraint \cite{BDTT}. For example, the discreteness of the volume operator in the kinematical Hilbert space of loop quantum gravity does not necessarily entail the discreteness of the volume operator once the quantum constraint is imposed.  Note that our result independently confirms and strengthens these previous results. We show that the ``nonpredictability'' is due to arbitrariness in the choice of internal clock and that there might be, in fact, many different spectra for a unique physical observable.

\section{Quantum mechanics regained}\label{VI}
In what follows, we show how to reconcile the proposed framework of quantum mechanics in internal clocks with quantum mechanics based on a fixed external time, which works so well in physicists' laboratories. For this purpose we will consider a system made of two free particles and equipped with some initial internal clock. The particles represent, respectively, a laboratory system and the environment.  We assume that states of the particle representing the environment and the initial clock provide a pool of new admissible clocks. The fact that we use for the purpose of the following analysis two particles instead of a more complex system is completely irrelevant. However, our analysis is based on two essential assumptions, which, in our minds, correspond to the real situation of the laboratory systems and their environments. We introduce them below.

Let us consider a system of two freely moving particles, denoted by P1 and P2, and a clock $t$. The canonical formalism consists of the phase space basic variables and an internal clock $(q_1,p_1,q_2,p_2)\in\mathbb{R}^4$. The dynamics of the total system is encoded in the contact form,
\begin{align}\label{ham12}\begin{split}
\omega_C =\omega_{C,1}+ & \omega_{C,2},\\
\omega_{C,i}=\ud q_i\ud p_i-\ud t\ud H_i,  ~~H_i & = \frac{p_i^2}{2},~~i=1,2.\end{split}
\end{align}
The canonical variables describe the positions and momenta of the particles. We denote the corresponding symplectic forms by 
$$\omega_i=\omega_{C,i}|_{t}.$$
The Hamilton equations read:
\begin{align}\label{HEP1}
\frac{\ud q_1}{\ud t}=-\omega_1^{-1}(q_1,H_1),~~\frac{\ud p_1}{\ud t}=-\omega_1^{-1}(p_1,H_1),
\end{align}
and
\begin{align}\label{HEP2}
\frac{\ud q_2}{\ud t}=-\omega_2^{-1}(q_2,H_2),~~\frac{\ud p_2}{\ud t}=-\omega_2^{-1}(p_2,H_2).
\end{align}
Now, we make the key assumptions:
\begin{description}
\item[1] ~~Particle P1 is quantum, and the canonical description presented above forms the underlying classical limit, whereas the particle P2 is described classically.
\item[2] ~~Particles P1 and P2 do not interact, which is encoded in Eq. (\ref{ham12}), and thus, each energy, $H_1$ and $H_2$, is conserved separately.
\end{description}

Let us consider a clock transformation that involves only external degrees of freedom to particle P1, that is, a transformation that depends on the states of particle P2,
\begin{align}\label{ctp1p2}
t\mapsto \bar{t}=t+D(q_2,p_2).
\end{align}
The respective pseudocanonical transformation reads
\begin{align}\label{PSq1}
\bar{q}_1=q_1-p_1D(q_2,p_2),~~\bar{p}_1=p_1,\\\label{PSq2}
\bar{q}_2=q_2-p_2D(q_2,p_2),~~\bar{p}_2=p_2,
\end{align}
in accordance with the previous sections, in particular with Eq. (\ref{psP}). We note that $D(q_2,p_2)$ depends only on the state of particle P2, the dynamics of which is classical and completely independent of particle P1 as is apparent from Eq. (\ref{HEP2}). As a result, from the point of view of the dynamics of particle P1, 
\begin{align}D(q_2(t),p_2(t))=\Delta(t)\end{align}
is an external, time-dependent parameter, and the transformation (\ref{PSq1}) is canonical. Why? It suffices to notice that from the point of view of the particle P1 the clock transformation (\ref{ctp1p2}) can be rewritten as
\begin{align}\label{ctd}
t\mapsto \bar{t}=t+\Delta(t),
\end{align}
and hence, it does not modify the constant time surfaces but merely relabels them. As a result, the symplectic form understood as a restriction of the respective contact form to constant time surfaces (\ref{symcon}),
\begin{align}\bar{\omega}_1=\omega_{C,1}|_{\bar{t}}=\omega_{C,1}|_{t+\Delta(t)}=\omega_1,\end{align}
remains unchanged.
Obviously, this cannot be said of particle P2 and upon the transformation (\ref{PSq2})
\begin{align}\bar{\omega}_2\neq\omega_2.\end{align}

Let us canonically quantize particle P1,
\begin{align}\label{qp1}q_1\mapsto Q,~~p_1\mapsto P,~~H_1\mapsto \frac12P^2,\end{align}
and the dynamics of the entire system is given by the Hamilton equations (\ref{HEP2}) combined with the Schr\"odinger equation,
\begin{align}i\frac{\partial |\psi\rangle}{\partial t}=\frac12P^2|\psi\rangle,~~|\psi\rangle\in\mathcal{H}\end{align} 
and occurs in the tensor product of one-particle phase space and the Hilbert space, $\mathbb{R}^2\otimes \mathcal{H}$.

From the pseudocanonical transformation (\ref{PSq1}), one finds that the observable $q_1$ in the phase space variables $\bar{q}_1,\bar{p}_1$ reads
\begin{align}q_1=\bar{q}_1-\Delta(t)\bar{p}_1.\end{align}
As discussed in Sec. \ref{IV}, the respective quantisation in $\bar{t}$ reads
\begin{align}\bar{q}_1\mapsto Q,~~\bar{p}_1\mapsto P,\end{align}
and hence the operator corresponding to the observable $q_1$ in the clock $\bar{t}$ reads
\begin{align}q_1\mapsto Q-\Delta(t)P,\end{align}
which, in comparison with Eq. (\ref{qp1}), shows that the unique observable $q_1$ (the position of particle P1) is in fact assigned a different operator in the two different clocks. There exists a unitary transformation $U$ corresponding to (\ref{PSq1}), which relates the basic operators
\begin{align}UQU^{\dagger}=Q-\Delta(t)P,~~UPU^{\dagger}=P,\end{align}
which implies
\begin{align}U=e^{-\frac{i\Delta}{2}P^2}.\end{align}
In fact, $U$ corresponds to a shift in time by $-\Delta$, as the quantum Hamiltonian reads $\hat{H}_1=\frac12P^2$. 

Suppose we use the position operator of particle P1 to assign a wave function to a state vector $|\psi\rangle\in\mathcal{H}$ in $t$:
\begin{align}\psi(q_1)=\langle q_1|\psi\rangle,~~Q|q_1\rangle=q_1|q_1\rangle.\end{align}
Then making the same decomposition of $|\psi\rangle\in\mathcal{H}$ in $\bar{t}$ gives
\begin{align}\varphi(q_1)=\langle q_1|U^{\dagger}|\psi\rangle,~~(Q-\Delta(t)P)U|q_1\rangle=q_1U|q_1\rangle,\end{align}
which is another wave function since the position operator is different now. In fact, they are related by $U$,
\begin{align}\psi(q_1)=U\varphi(q_1)\end{align}
as $U$ relates the eigenvectors corresponding to the same value in the position spectrum, $q_1$. Therefore, observers associated with different clocks must interpret a given state $|\psi\rangle\in\mathcal{H}$ in the same way except that they make their interpretation at different moments, shifted by $\Delta(t)$, since they differently label the same constant time surfaces. Thus, there is no clock effect. 

The evolution of wave functions given by the respective Schr\"odinger equations are physically identical in both clocks as
\begin{align}i\partial_t\psi(q_1)=\hat{H}\psi(q_1)\end{align}
implies
\begin{align}U^{\dagger}i\partial_tU\varphi(q_1)=\hat{H}\varphi(q_1),\end{align}
which expressed explicitly reads
\begin{align}\frac{i}{1+\partial_t\Delta(t)}\partial_t\varphi(q_1)=\hat{H}\varphi(q_1),\end{align}
and is equivalent to
\begin{align}i\partial_{\bar{t}}\varphi(q_1)=\hat{H}\varphi(q_1)\end{align} by the virtue of the clock transformation (\ref{ctd}). Therefore, the described clock transformation represents merely a change of units and origin of time with no consequences for the quantum theory.

We conclude the present section by encapsulating our finding as follows: given a quantum system, clock transformations that involve only external and classical degrees of freedom merely change the units of time in its quantum description. This kind of transformations explains why unique Schr\"odinger equations apply to laboratory setups that make a clear division between examined quantum systems (P1) and classical environments (P2). Therefore, quantum mechanics in internal clocks contains quantum mechanics in a fixed time as a special case.

\section{Conclusions}\label{VII}

Quantum mechanics is based on an external and fixed parameter called time. According to general relativity no such entity exists. Thus, the evolution of gravitational systems is described in terms of internal degrees of freedom. In this work we have proposed a reformulation of quantum mechanics in such a way as to remove the absolute time from its formalism and replace it with an arbitrarily chosen internal degree of freedom, the internal clock.

We arrive at our formulation by first extending the symmetry of the canonical formalism of classical mechanics. Namely, canonical transformations are replaced by pseudocanonical transformations which include the former as a normal subgroup and which allow us to switch from one internal clock to another. The canonical formalisms equipped with different internal clocks are physically completely equivalent.

Next, we lift the extended symmetry to the quantum level. Quantizations of canonical formalisms in all internal clocks satisfy the basic postulate of the new formulation: all Dirac observables are represented by the same quantum operators in a fixed Hilbert space irrespectively of the choice of clock. Therefore, any state of a quantum system is unambiguously represented by a vector in the fixed Hilbert space, which can be unambiguously expressed by a wave function in any Dirac observable representation. Moreover, the motion of state vectors in the Hilbert space is unique as it is determined by the Schr\"odinger equation based on an unambiguously quantized Dirac observable, the Hamiltonian.

Then, we show that the choice of internal clock influences the dynamical properties of quantum states. This is so because dynamical observables are represented by different quantum operators for different internal clocks. Hence, for a fixed vector in the Hilbert space, the form of the wave function in any dynamical observable representation depends on the choice of internal clock. As we show in the simple example of a free particle, the spectrum of the position operator can even have a different character that depends on the chosen internal clock.

The described nonequivalence of different internal clock frames for a given quantum system is a distinctive and inevitable property of the internal clock formulation of quantum mechanics. We call it the clock effect. 

Finally, the new formulation contains the usual formulation of quantum mechanics as a special case. When we study a quantum system in a laboratory, we separate it from its environment that is classical. The environment provides a supply of internal clocks that do not influence the quantum observables for the laboratory system. Therefore, the environment plays a role of the external and fixed time, which can be assumed in the formalism. From this, we conclude that our reformulation is completely consistent with the ordinary formalism and automatically reproduces all its successful results. On the other hand, it seems to be more appealing theoretically.

The most important gain from our reformulation of quantum mechanics is that it can now be applied to gravitational systems. In comparison with the ordinary quantum mechanics, the presented formalism places even more constraints on the classical interpretation of physical systems. It implies for the early Universe multiple ordinary quantum scenarios of the birth of the Universe. At the same time, it predicts that, as the universe expands and classical environments are formed, its quantum dynamical properties converge. Finally, the Universe at large scales approaches a completely classical and unique state. For application of the internal clock formulation of quantum mechanics to a cosmological system, see, e.g. \cite{M2}.

\section*{Acknowledgments} 
This work was supported by Narodowe Centrum Nauki with Decision No. DEC-2013/09/D/ST2/03714.

\section*{Appendix: Solution to the eigenvalue problem}
Our discussion of the spectral properties of the operators considered in Sec. \ref{eg2} is based on the von Neumann theory of unbounded self-adjoint operators on Hilbert space described, e.g., in \cite{reedsimon}. In Sec. \ref{eg2}, we quantize the position of a particle on the real line, denoted by $q$, with respect to unusual clocks related to the initial one as $\bar{t}=t+D$, where $D$ is given in Eq. (\ref{delayP}). The respective position operator is given in Eq. (\ref{new_operators}). Let us introduce an auxiliary function $h(p)=1+pf(p)$ and give the operator (\ref{new_operators}) the following form
\begin{equation}\label{appq}
\frac{1}{2}\left( i \frac{\partial}{\partial p} \cdot h(p)+h(p) \cdot i \frac{\partial}{\partial p} \right).
\end{equation}
The above form can be simplified by means of unitary transformations. The first one reads
\begin{equation}
\begin{split}
L^2(\mathbb{R},dp)&\mapsto L^2(\mathbb{R},h^{-1}(p)\ud p),\\
\psi(p) &\mapsto \varphi(p)=h^{1/2}(p)\psi(p),
\end{split}
\end{equation}
and hence,
\begin{equation}
i \frac{\partial}{\partial p} \psi(p)\mapsto h^{1/2}  (p) i \frac{\partial}{\partial p} h^{-1/2}(p) \varphi(p).
\end{equation}
The second one reads 
\begin{equation}
\begin{split}
L^2(\mathbb{R},h^{-1}\ud p)&\mapsto L^2(\mathbb{X}, \ud z),\\
\varphi(p) &\mapsto \phi(z)=\varphi(p(z)),
\end{split}
\end{equation}
where $\frac{\ud p}{\ud z}=h(p)$ and
\begin{equation}
i \frac{\partial}{\partial p}\varphi(p) \mapsto \frac{\ud z}{\ud p} \cdot \frac{\partial}{\partial z}\phi(z).
\end{equation}
The application of the above unitarity transformations to the position operator (\ref{appq}) leads to
\begin{equation}
\frac{1}{2}\left( i \frac{\partial}{\partial p} \cdot h(p)+ h(p)\cdot i \frac{\partial}{\partial p} \right)\psi(p) \mapsto i \frac{\partial}{\partial z}\phi(z),
\end{equation}
where
\begin{equation}
\phi(z)=h^{1/2}\psi(p(z))\in L^2(\mathbb{X}, \ud z),
\end{equation}
that is, the position operator is simplified to the usual form of the ``momentum operator" on the space $\mathbb{X}$. The solution of the eigenvalue problem is straightforward; the only issue is the space $\mathbb{X}$ itself and we distinguish three cases:

\vspace{3 mm}
\textit{1.} $\mathbb{X}=\mathbb{R}$ if the integrals $\int_{0}^{\pm\infty} \frac{\ud p}{h(p)} $ are infinite.

\textit{2.} $\mathbb{X}=\mathbb{R}_+$ if one of the integrals $\int_{0}^{\pm\infty} \frac{\ud p}{h(p)}$ is finite. 

\textit{3.} $\mathbb{X}=\mathbb{I}$ (where $\mathbb{I}$ stands for an interval) if the integral $~~\int_{-\infty}^{\infty} \frac{\ud p}{h(p)} $ is finite. 
\vspace{3 mm}

It is well known that for case \textit{1} the momentum operator is self-adjoint and has a continuous real spectrum; for case \textit{2}, the momentum operator has no self-adjoint extension (and we omit this case); for case \textit{3}, the momentum operator has many inequivalent self-adjoint extensions and each of them possesses an unbounded and discrete spectrum.

In Sec. \ref{eg2}, we consider two cases, A) the discrete one $h(p)=p^2+1$, and B) the continuous one
$h(p)=(3 p^2 + 1)^{-1}$. Both satisfy the monotonicity condition (\ref{mono}), which now reads
\begin{equation}
h(p)>0.
\end{equation}

\end{document}